\documentclass[aps,prb, twocolumn, showpacs,preprintnumbers,amsmath,amssymb,superscriptaddress]{revtex4-1}
         
\usepackage{graphicx}
\usepackage{dcolumn}
\usepackage{bm}
\usepackage{longtable}
\usepackage{color}
\usepackage{textcomp}
\usepackage{hyperref}  
\usepackage{gensymb}
 
\begin{document}     

\preprint{PREPRINT (\today)}

\newcommand{\YBCO}{YBa$_2$Cu$_3$O$_7$}
\newcommand{\Tc}{$T_{\rm c}$}
 
\title{Vortex lock-in transition coinciding with the 3D to 2D crossover in \YBCO}
    
\author{S.~Bosma} 
\email{sbosma@physik.uzh.ch}
\affiliation{Physik-Institut der Universit\"at Z\"urich, Winterthurerstrasse 190, CH-8057 Z\"urich, Switzerland}

\author{S.~Weyeneth}
\affiliation{Physik-Institut der Universit\"at Z\"urich, Winterthurerstrasse 190, CH-8057 Z\"urich, Switzerland}

\author{R.~Puzniak}
\affiliation{Institute of Physics, Polish Academy of Sciences, Aleja Lotnik\'ow 32/46, PL-02-668 Warsaw, Poland}

\author{A.~Erb}
\affiliation{Walther Meissner Institut, Bayerische Akademie der Wissenschaften, D-85748 Garching, Germany}

\author{H.~Keller}
\affiliation{Physik-Institut der Universit\"at Z\"urich, Winterthurerstrasse 190, CH-8057 Z\"urich, Switzerland}
 
\begin{abstract}

A vortex lock-in transition was directly detected by torque magnetometry in an overdoped \YBCO\ single crystal of low anisotropy ($\approx 7$). The locked-in state was observed below the 3D to 2D crossover temperature $T_{\rm cr} = 76~K$, independently of extrinsic pinning effects thanks to a high quality clean crystal and the use of a vortex shaking technique. The shape of the torque signal as a function of the angle between the applied magnetic field and the crystallographic $c$-axis is in very good agreement with the model developped by Feinberg and Ettouhami [Int. J. Mod. Phys. B {\bf 7}, 2085 (1993)] for quasi-2D superconductors, despite the low anisotropy  of the material.

\end{abstract}
  
\pacs{74.20.De; 74.25.Ha; 74.72.-h}
 
\maketitle

\section{Introduction}

Dimensionality is essential to understand the behavior of vortices in layered cuprate superconductors.  A three-dimensional (3D) to two-dimensional (2D) crossover takes place when the superconducting coherence length along the $c$-axis $\xi_c$ becomes smaller than the distance $s$ between the planes supporting superconductivity.\cite{Blatter1994} Since $\xi_c$ decreases with decreasing temperature, it will in many cases become smaller than $s$ below some temperature. The Lawrence-Donniach model should then be used to describe superconductivity.\cite{Lawrence1972} The temperature $T_{\rm cr}$ at which this crossover happens is such that $\xi_c(T_{\rm cr})=\xi_{c, 0}/\sqrt{1-T_{\rm cr}/T_{\rm c}}=s$, where $\xi_{c, 0} = \xi_c(T =~0~{\rm K})$ and $T_{\rm c}$ is the superconductor critical temperature. For \YBCO, taking $s \simeq 0.8$ nm and $\xi_{c, 0} \simeq 0.3$ nm, one gets $T_{\rm cr} \simeq 76$ K.  \\
\indent In the 2D regime, when the applied magnetic field direction is nearly parallel to the $ab$-plane, a lock-in transition may take place.\cite{Feinberg1990} In this case, the vortex cores are confined between the superconducting layers, even though the field is not aligned with these layers. This minimizes condensation energy at the cost of magnetic energy coming from the misalignement of vortices and field, since the cores do not cross the layers anymore. This is also known as intrinsic pinning, since it locks the vortices independently of (extrinsic) impurities. \\
\indent In high anisotropy materials lock-in studies,\cite{Steinmeyer1994, Janossy1995, Okram2001, Zehetmayer2005} the 2D character is so strong that the lock-in is present almost up to \Tc. However, the vicinity of the superconducting transition makes it difficult to observe the lock-in onset. Low anisotropy cuprates like \YBCO\ or YBa$_2$Cu$_4$O$_8$ are more suited for this purpose. The lock-in was observed in \YBCO\ by torque magnetometry,\cite{Farrell1990, Kortyka2010} bulk resistivity measurements,\cite{Kwok1991, Gordeev2000} and AC transport in thin films.\cite{Doyle1993}  The lock-in was also observed in  various other layered superconductors.\cite{Vermeer1991, Mansky1993, Kolesnik1996, Bugoslavsky1997, Avila2001, Khene2004, Kohout2007a} It may be difficult to distinguish between pinning and lock-in effects (see for example Ref.~\onlinecite{Kortyka2010}). Besides, a large irreversibility due to extrinsic pinning effects may hide the appearance of the lock-in transition: in Ref.~\onlinecite{Zech1996a}, the lock-in transition is identified much below the 3D to 2D crossover temperature $T_{\rm cr}$. In this work, the appearance of the vortex lock-in is clearly observed by torque magnetometry at $T_{\rm cr}$ in a clean overdoped \YBCO\ single crystal. \\
\indent A review of lock-in theoretical models is given in Ref.~\onlinecite{Blatter1994} (p. 1286). The most relevant models for this work are presented in Refs.~\onlinecite{Bulaevskii1991, Feinberg1993}. The lock-in angle corresponds to the angle between the applied magnetic field and the crystallographic $c$-axis at which the lock-in appears. This angle is the crucial parameter turning the lock-in on and off. Previous experiments on various cuprate superconductors were in agreement with the theory whenever data accuracy made the comparison possible, but the data in the case of \YBCO\ were rather sparse. In this work, we present a detailed study of the field and temperature dependence of the lock-in effect in a low anisotropy cuprate superconductor. We note a very good qualitative agreement with the behavior described in Ref.~\onlinecite{Feinberg1993}, although the field dependence of the lock-in angle seems unconventionnal.  The field $H$ is chosen in the London domain $H_{\rm c1} << H << H_{\rm c2}$, where $H_{\rm c1}$ and $H_{\rm c2}$ are the lower and upper critical fields. This excludes the interference of other phenomena like vortex lattice melting or glass behavior. The temperature range has a lower bound of 60 K, because irreversibility renders the data unreliable below this temperature; the torque ceases to conform to the model described in Ref.~\onlinecite{Feinberg1993}.      
 
\section{Torque measurements}

\begin{figure} 
\centering
\includegraphics[width=1\linewidth]{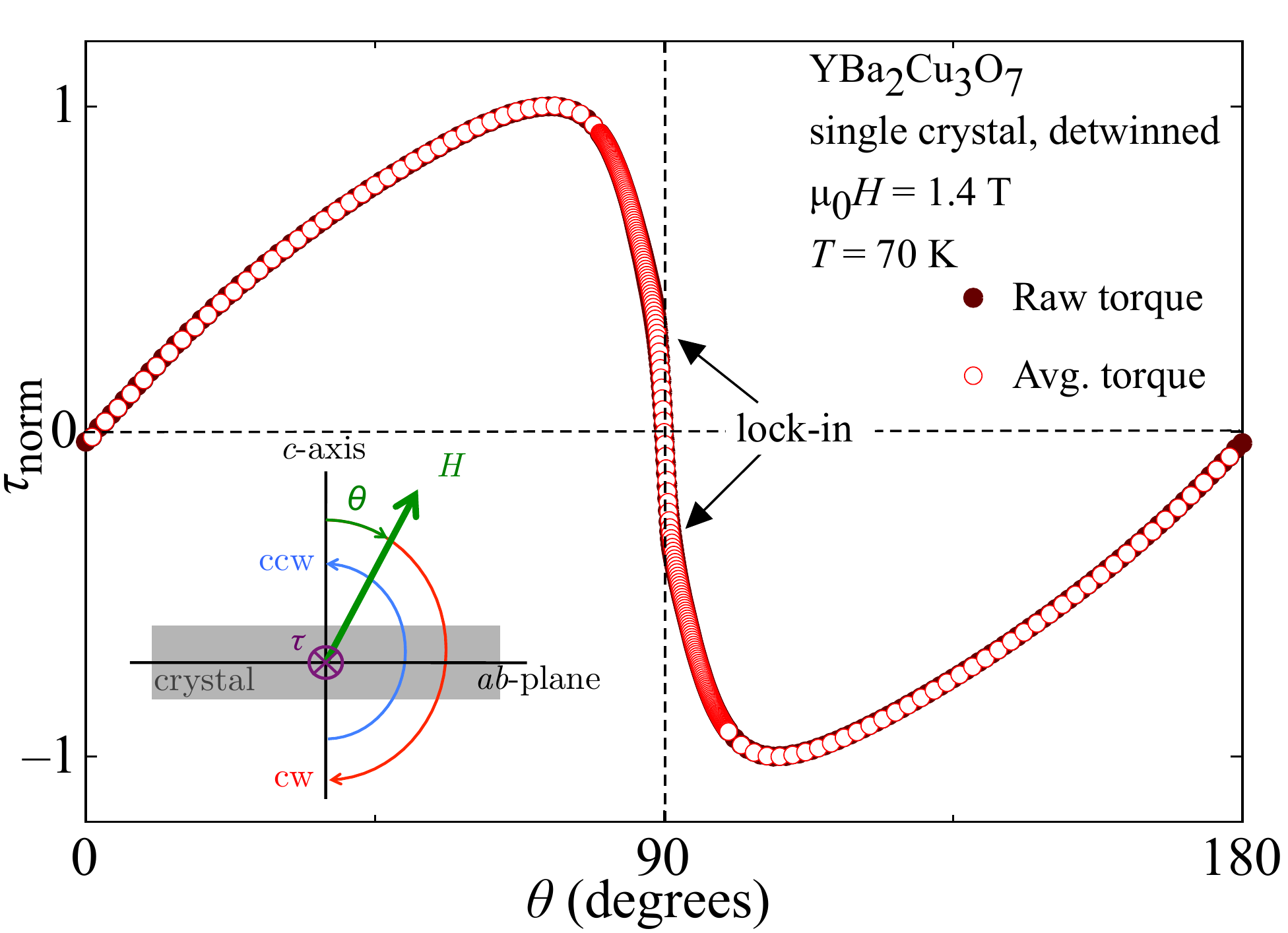}
\caption{(Color online) Normalized torque $\tau/\tau_{\rm max}$ of detwinned single crystal \YBCO\ in the full angular range at 70 K and 1.4 T. The irreversibility is very small; at this scale, the average of the clockwise (CW) and counter-clockwise (CCW) branches of the torque (red open circles) overlaps the raw unaveraged signal (brown closed circles). Inset: Field orientation during CW and CCW measurements.}
\label{fullAngle}
\end{figure}

\indent The growth procedure and detwinning of the high-quality superconducting \YBCO\ single crystal used in this experiment is described in Ref.~\onlinecite{Erb1999}. The dimensions of the platelet crystal are 130$\times$160$\times$50 $\mu$m$^3$, and $T_{\rm c}~\simeq~88$ K. Magnetic torque investigations were carried out using a home-made magnetic torque sensor.\cite{Kohout2007} The sample is attached to a platform hanging on piezoresistive legs. When a magnetic field is applied on an anisotropic superconductor, the misalignement between field and diamagnetic moment results in a torque. This bends the legs, thus giving rise to a measureable electric signal proportional to the magnetic torque. For a uniaxial superconductor, the angular dependence of the magnetic torque $\vec{\tau}=\vec{m} \times \mu_0 \vec{H}$  (where $\vec{m}$ is the sample magnetic moment) in the London approximation ($H_{c1} << H << H_{c2}$) can be written as\cite{Kogan1988}

\begin{equation}
\label{kogan}
\tau(\theta,H)=AH\frac{\sin(2\theta)}{\epsilon(\theta)}\ln\left(\frac{\eta H_{\rm c2}^{||c}}{\epsilon(\theta)H}\right),
\end{equation}
where $\theta$ is the angle between the applied magnetic field $H$ and the crystallographic $c$-axis, $\epsilon(\theta)=\sqrt{\cos^2(\theta)+\sin^2(\theta)/\gamma_\lambda^2}$ is the angular scaling function, $H_{\rm c2}^{||c}$ is the $c$-axis upper critical field, and $\eta$ is a dimensionless parameter of the order of unity. The anisotropy parameter $\gamma_\lambda=\lambda_c$/$\lambda_{ab}$ is the ratio of the out-of-plane and in-plane magnetic penetration depths; $A =  -V\Phi_0(1-1/\gamma_\lambda^2)/(16\pi\lambda_{ab}^2)$ ($V$ is the sample volume, $\Phi_0$ is the flux quantum) is independent of angle. This model is 3-dimensionnal.  \\
\indent A typical torque signal $\tau(\theta)$ of \YBCO\ measured in a magnetic field of 1.4 T at 70 K is shown in Fig.~\ref{fullAngle}. During a torque measurement, the field direction is swept clockwise (CW) from the $c$-axis (0$^\circ$) through the $ab$-plane (90$^\circ$) and the opposite direction of the $c$-axis (180$^\circ$), then swept back counter-clockwise (CCW), as shown in the inset of Fig.~\ref{fullAngle}. The CW and CCW branches of the raw unaveraged torque signal overlap when the torque is reversible. The torque signal is antisymetric with regards to the $ab$-plane, so the rest of the data are shown only in the angle range 0$^\circ$ to 90$^\circ$ for clarity.

\begin{figure} 
\centering
\includegraphics[width=1\linewidth]{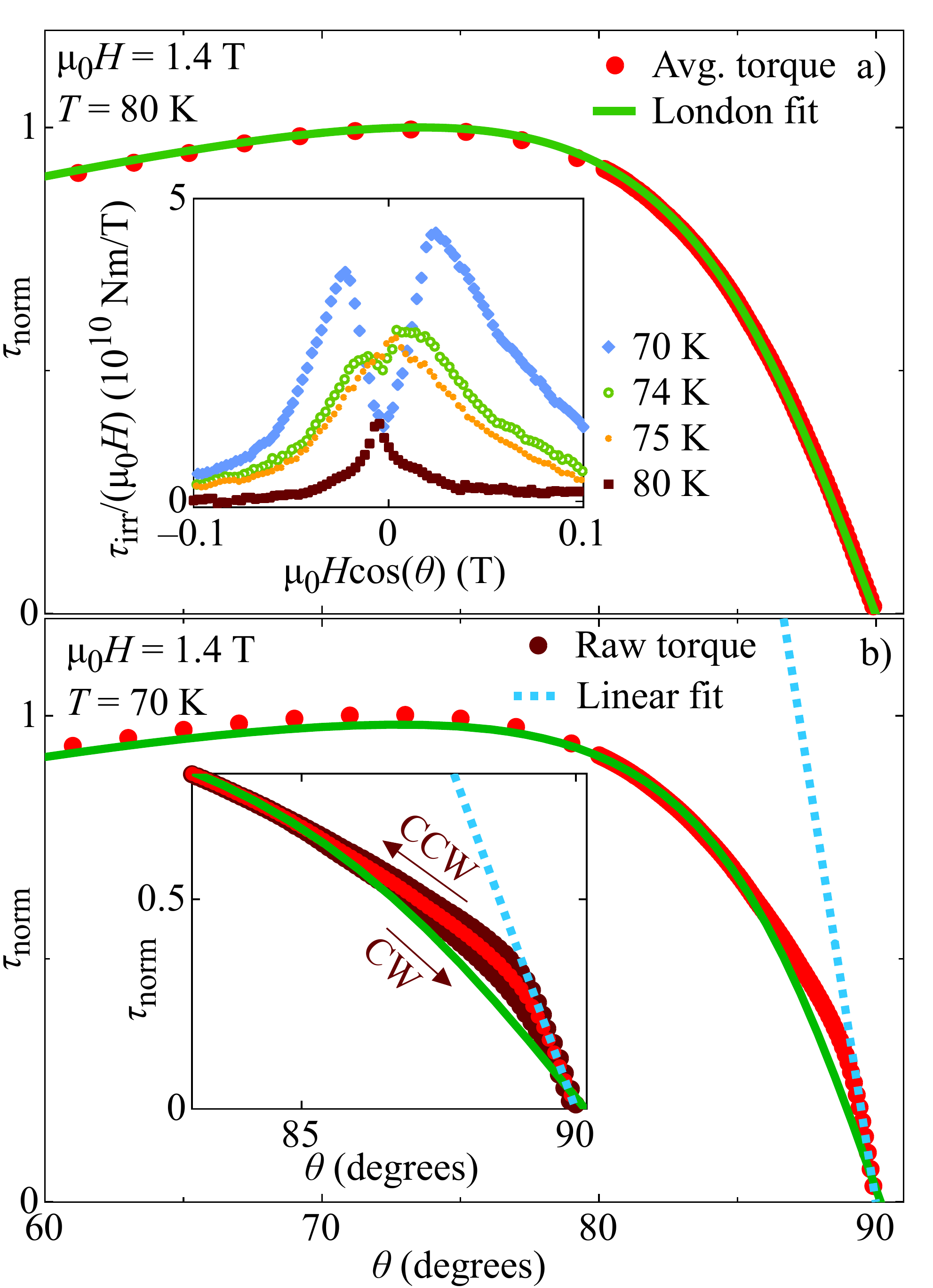}
\caption{(Color online)  a)  Normalized torque of detwinned single crystal \YBCO\ as a function of angle at 80 K and 1.4 T. The torque follows the London dependence at all angles. The green line is a fit of Eq.~(\ref{kogan}) to the data.  The light red points represent an average of the CW and CCW angle measurements. Inset: Irreversible component of the torque $\tau_{\rm irr} = (\tau_{\rm CW} - \tau_{\rm CCW})/2$ at various temperatures at 1.4 T.  b) Normalized torque as a function of angle at 70 K and 1.4 T. The dotted blue line is a linear fit of the lock-in region. Inset: Close-up around the $ab$-plane. The dark red curve is the raw torque. The arows indicate the field sweep direction for each branch. }
\label{overview}
\end{figure}

\begin{figure} 
\centering
\includegraphics[width=1\linewidth]{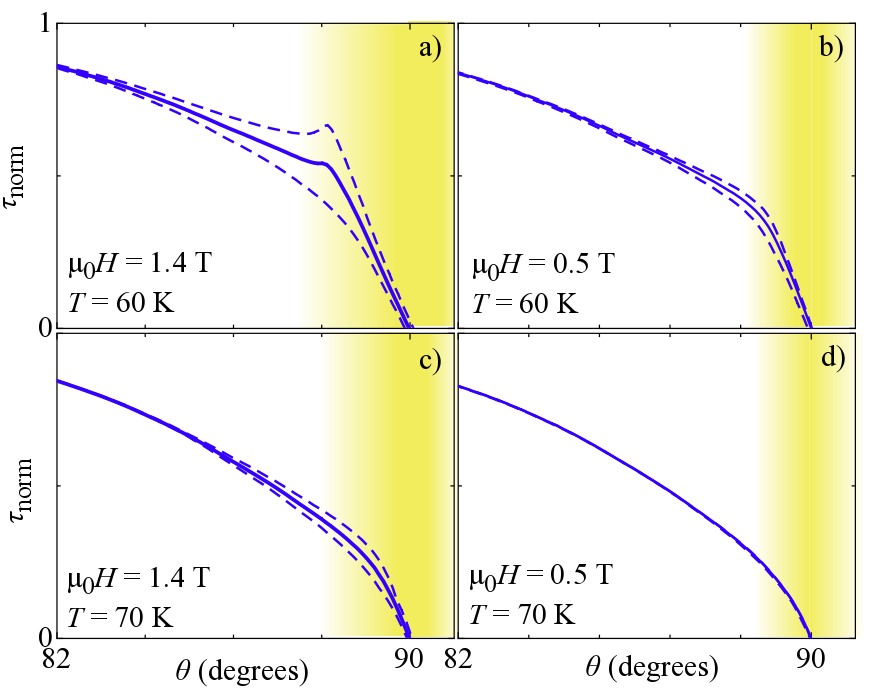}
\caption{(Color online) Normalized torque of detwinned single crystal \YBCO\ as a function of angle. The dashed line is the raw irreversible data, the solid line is the average between the CW and CCW measurements. The irreversibility is limited to the angle range where the torque does not follow the London model above 80$^\circ$. The yellow gradient zone underlines the lock-in region and the transition zone. The lock-in transition is not sharp at low fields and high temperatures.  a) 1.4 T and 60 K. At these low temperature, the torque does not follow the shape described in Ref.~\onlinecite{Feinberg1993} b) 0.5 T and 60 K. This shape is predicted in Ref.~\onlinecite{Feinberg1993}. c) 1.4 T and 70 K. d) 0.5 T and 70 K. The shaking is optimal: the signal is fully reversible.}
\label{irrev}
\end{figure}

\indent Figure~\ref{overview} shows a  torque measurement evidencing the lock-in effect; a deviation from Eq.~(\ref{kogan}) can be seen at low temperatures close to the $ab$-plane for $\theta \simeq 85^\circ$. This corresponds to a staircase configuration\cite{Feinberg1993} of the vortices. The order parameter inside a vortex core is not suppressed between the layers, because the circulating currents are Josephson currents and not superconducting currents. The order parameter is only suppressed within the superconducting layers. The vortex consists of 2D cores in the layers, linked by Josephson cores between the layers. When the vortices are tilted enough that the Josephson coherence length is smaller than the distance between two consecutive vortex cores, the vortex line takes a staircase shape. The physics stays 3D on large scales, but the free energy deviates from the 3D London model.\cite{Feinberg1992} For $\theta \simeq 87^\circ$ the lock-in starts: the torque becomes linear and changes slope. This shape of the averaged angular dependent torque is identical to the prediction of the model presented in Ref.~\onlinecite{Feinberg1993} for quasi-2D superconductors. This similarity is striking, as the anisotropy of our \YBCO\ crystal is around 7,\cite{Bosma2011} which would not qualify as quasi-2D. The models presented in Ref.~\onlinecite{Feinberg1993} relate to anisotropies around 50, as expected in La$_{2-x}$Sr$_x$CuO$_4$ for example. \\
\indent The torque data exhibits an  angular irreversibility between the clockwise (CW) and counter-clockwise (CCW) branches. Such irreversible signals are usually due to vortex pinning. In this work, the so-called vortex-shaking technique\cite{Willemin1998} was applied to reduce irreversibility. This was done by applying a small AC field orthogonal to the main field $H$ in order to enhance the vortex relaxation towards thermodynamic equilibrium. The irreversible part of the torque $\tau_{\rm irr} = (\tau_{\rm CW} - \tau_{\rm CCW})/2$ is shown in the inset of Fig.~\ref{overview}a. The shape of $\tau_{\rm irr}$ changes when the lock-in appears. The double peak shape of $\tau_{\rm irr}$ is characteristic\cite{Zech1996a} of the lock-in state. This confirms the lock-in transition temperature around 75 K. The small residual irreversibility as seen on the averaged torque could not be hiding a small higher temperature lock-in signal, because lock-in shows up as well in the shape of $\tau_{\rm irr}$. Besides, the difference between the London fit and the averaged torque is too large to be an artifact of irreversibility (Fig.~\ref{overview}b inset). \\
\indent Figure~\ref{irrev} shows the angular torque at various temperatures and fields. A small residual irreversibility is visible close to the $ab$-plane, at the same angles where lock-in takes place. This irreversibility decreases with increasing temperature, as expected for vortex pinning. It also decreases with decreasing field, as observed in YBa$_2$Cu$_4$O$_8$,\cite{Zech1996a} but contrary to what was observed in \YBCO.\cite{Farrell1990} The dependence of the pinning forces on field depends on the crystal quality and field range of the experiment, which may explain this different field behavior. In a clean crystal like the one used in this work, the only source of pinning is the layered structure; the appearance of irreversibility is thus a supplementary indication of the onset of lock-in.
 
\begin{figure} 
\centering
\includegraphics[width=1\linewidth]{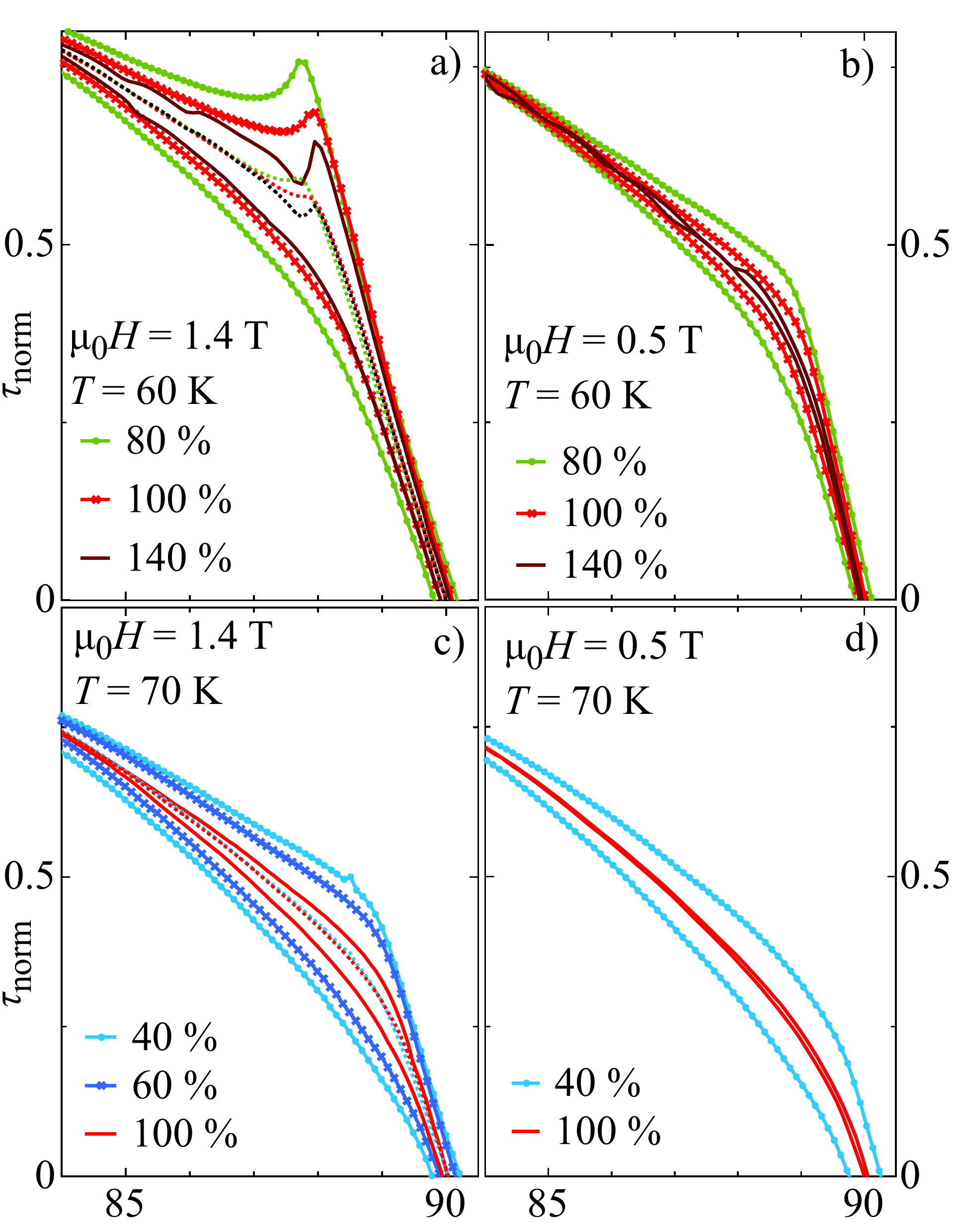}
\caption{(Color online) Effect of shaking power on the normalized angular torque of detwinned single crystal \YBCO. $100 \%$ represents the maximal stable shaking field. Above this value, the shaking power is difficult to keep at a constant level for the whole duration of the measurement due to equipement limitations. a) High field, low temperature: the averaged torque signal (dotted lines) depends on shaking power. b) Low field, low temperature: the shaked torque can be made almost reversible, and the averaged torque (not shown for clarity) stays constant with shaking power.  c) High field, high temperature: the averaged torque signal (dotted lines) does not depend on shaking power. d) Low field, high temperature: the shaking is optimally efficient in this field-temperature domain, {\it i.e.} the shaked torque can be made reversible.}
\label{shakingEffect}
\end{figure}

\indent In order to investigate the effect of vortex shaking on the lock-in phenomenon, we studied different shaking field amplitudes at various static fields and temperatures. At low temperature and high fields (Fig.~\ref{shakingEffect}a), the shaking is not sufficient to ensure reliable measurements, since the averaged data depend on the shaking power. The linear lock-in zone is reduced by increasing shaking power. The shaking efficiency limit is reached when a small peak appears at the limit of the lock-in domain. The peak feature in the CCW branch of the torque is characteristic of lock-in observed in conjunction with extrinsic pinning.\cite{Kortyka2010} This usually masks the lock-in effect in lower quality crystals. In this work the peak appears only if the extrinsic pinning becomes too large to be suppressed at low temperatures. At low temperature and low field (Fig.~\ref{shakingEffect}b), and high temperature and high field (Fig.~\ref{shakingEffect}c), the shaking power is sufficient to get stable data. All the averaged torque signals for the various shaking powers are the same. At low fields and high temperatures (Fig.~\ref{shakingEffect}d), the shaking is even sufficient to get fully reversible data. We consider that the lock-in properties are reliably measured if increasing the shaking power does not change the shape of the averaged torque. 

\section{Discussion}

The analysis was done on the average of the CW and CCW data, since the deviation of the averaged data from the London model is larger than the irreversibility, as also reported in Ref.~\onlinecite{Farrell1990}. The lock-in angle is often viewed as the angle at which the perpendicular component of the field goes below the lower critical field along the $c$-axis. Field penetration across the layers is then impossible, effectively locking the vortices in the $ab$-planes. In this model, the lock-in angle $\theta_{\rm lock}$ should be such that $H_{\rm lock}^{|| c} = H \cos(\theta_{\rm lock})$ equals $H_{\rm c1}^{|| c}$, so $\theta_{\rm lock}$ should decrease with increasing field. This evolution was observed in high anisotropy cuprates,\cite{Zehetmayer2005} and a similar dependence was derived for \YBCO,\cite{Feinberg1990, Bulaevskii1991} although not confirmed by experiments in this material. However, Ref.~\onlinecite{Vulcanescu1994} reports in La$_{2-x}$Sr$_x$CuO$_4$ ($x$ = 0.075) a transverse lock-in field $H_{\rm lock}^{ || c}$ which is different from $H_{\rm c1}^{|| c}$. It is thus possible that this simple $H_{\rm c1}$ picture holds only for high-anisotropy compounds, and that the lock-in angle is not necessarily inversely proportionnal to the field.

\begin{figure} 
\centering
\includegraphics[width=1\linewidth]{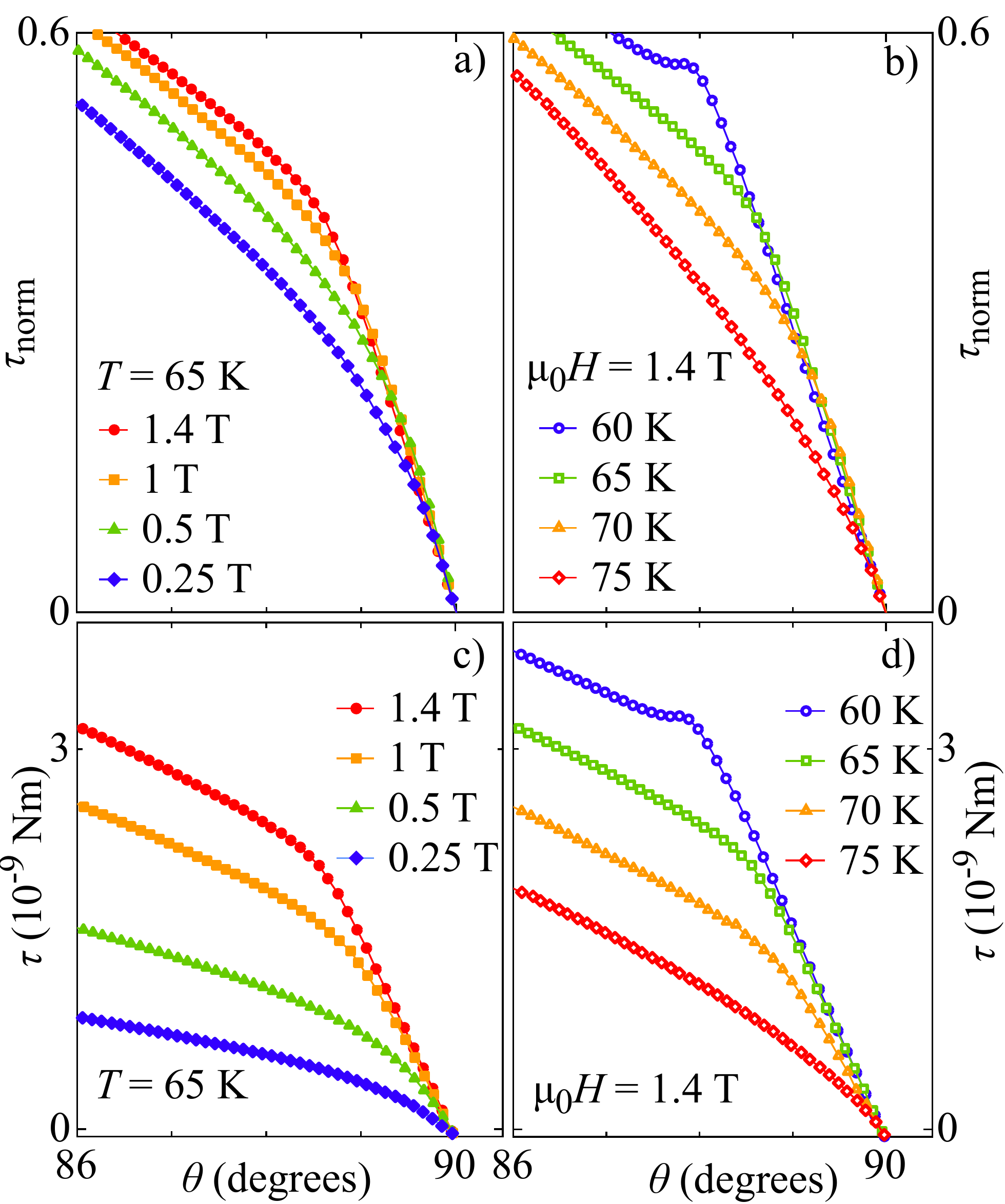}
\caption{(Color online) a) Normalized torque of detwinned single crystal \YBCO\ as a function of angle at 65 K for various fields. The lock-in angular domain and the lock-in amplitude are smaller at low fields. b) Normalized torque as a function of angle at 1.4 T for various temperatures. The lock-in angular domain and the lock-in amplitude are larger at high temperatures. c) Torque  as a function of angle at 65 K for various fields. d) Torque as a function of angle at 1.4 T for various temperatures.}
\label{TH}
\end{figure}

The vortices direction is along the magnetic induction $\vec{B}$; $\vec{B}$ is therefore aligned with the planes in the lock-in state (but $\vec{H}$ is not). Since $\vec{B} = \mu_0(\vec{H}+\vec{M})$ and $M$ is small compared to $H$, $B$ can be approximated as the parallel component of $H$: $B = \mu_0H^{|| ab} = \mu_0H\cos(\theta)$. In that case, the torque $|\vec{\tau}| = |V \vec{M}\times \mu_0\vec{H}| = V\mu_0HB\sin(\theta)$ becomes:

\begin{equation} 
\label{lockinEq}
\tau_{\rm lock} = V\mu_0 H^2 \sin(\theta) \cos(\theta).
\end{equation} 

Figure~\ref{TH} shows the lock-in transition at various fields and temperatures. If we define the lock-in angle $\theta_{\rm lock}$ as the angle where the torque slope changes (moves away from the linear region), $\theta_{\rm lock}$ increases at low temperatures, but also at high fields. However, $\theta_{\rm lock}$ becomes more difficult to identify at higher temperatures and lower fields.  It is possible that the observed unconventional increase of $\theta_{\rm lock}$ at high fields is biased; since the transition is smoother at low fields, the field dependence of $\theta_{\rm lock}$ might be drowned in the large transition. $\theta_{\rm lock}$ can also be defined as the point at which the torque is no longer independent of temperature, and therefore not following Eq.~(\ref{lockinEq}). In that case, we can reliably identify the temperature dependence of $\theta_{\rm lock}$ from measurements at constant field. Figure~\ref{TH}c and d show the non normalized torque in the lock-in region; it appears that even though all curves merge around the $ab$-plane, the torque slope depends slightly on temperature, contrary to the prediction of Eq.~(\ref{lockinEq}). At low temperatures, the transition sharpness increases and the slope depends more weakly on temperature. Since the lock-in transition is not sharp at higher temperatures, the curvature of the torque that accompanies this transition may skew the linear region and change the slope dependence given by Eq.~(\ref{lockinEq}).

\section{Conclusion}

A lock-in transition was observed in a clean detwinned \YBCO\ single crystal at the 2D to 3D crossover temperature. Although the angular torque signal matches theoretical shapes, it seems difficult to qualitatively confirm a simple model of the lock-in. The lock-in angle domain decreases with increasing temperature, as expected for vortex pinning. Surprisingly, this domain also seems to increase with field in the studied field range (0 to 1.4 T), although this dependence may be an artifact of a broad lock-in transition. This unconventional behavior might be related to the low anisotropy of the compound, which prevents it from having a strong 2D behavior, even at low temperatures.

\section{Acknowledgements}

\indent Helpful discussions with D. Feinberg and V.~B. Geshkenbein are gratefully acknowledged. This work was supported by the Swiss National Science Foundation.


%

\end{document}